\documentclass[twocolumn,english,prb]{revtex4}
\usepackage[T1]{fontenc}
\usepackage[latin9]{inputenc}
\usepackage{babel}
\usepackage{amsmath}
\usepackage{amssymb}
\usepackage{graphicx}
\usepackage{esint}
\usepackage[unicode=true,
 bookmarks=true,bookmarksnumbered=true,bookmarksopen=false,
 breaklinks=false,pdfborder={0 0 1},backref=false,colorlinks=false]
 {hyperref}
\hypersetup{
 pdfauthor={Achilleas Lazarides}}

\makeatletter
\@ifundefined{textcolor}{}
{%
 \definecolor{BLACK}{gray}{0}
 \definecolor{WHITE}{gray}{1}
 \definecolor{RED}{rgb}{1,0,0}
 \definecolor{GREEN}{rgb}{0,1,0}
 \definecolor{BLUE}{rgb}{0,0,1}
 \definecolor{CYAN}{cmyk}{1,0,0,0}
 \definecolor{MAGENTA}{cmyk}{0,1,0,0}
 \definecolor{YELLOW}{cmyk}{0,0,1,0}
}

\usepackage{babel}
\usepackage{color}
\usepackage{xargs}[2008/03/08]
\usepackage{braket}
\usepackage{braket}

\@ifundefined{textcolor}{}{%
 \definecolor{BLACK}{gray}{0}
 \definecolor{WHITE}{gray}{1}
 \definecolor{RED}{rgb}{1,0,0}
 \definecolor{GREEN}{rgb}{0,1,0}
 \definecolor{BLUE}{rgb}{0,0,1}
 \definecolor{CYAN}{cmyk}{1,0,0,0}
 \definecolor{MAGENTA}{cmyk}{0,1,0,0}
 \definecolor{YELLOW}{cmyk}{0,0,1,0}
}

\makeatother

\begin{document}

\title{The fate of a discrete time crystal in an open system}

\author{Achilleas Lazarides$^{1}$ and Roderich Moessner$^{1}$}
\affiliation{$^{1}$ Max-Planck-Institut f\"ur Physik komplexer Systeme, 01187 Dresden, Germany}

\newcommand{\tred}[1]{{\color{red} {#1}}}
\newcommand{\tgreen}[1]{{\color{green} {#1}}}
\newcommand{\rhospinor}[2]{\mathbf{\tilde{\rho}}_{#1,#2}}

\begin{abstract}
  Following the recent realisation that periodically driven quantum
  matter can support new types of spatiotemporal order, now known as
  discrete time crystals (DTCs), we consider the stability of this
  phenomenon. Motivated by its conceptual importance as well as its
  experimental relevance we consider the effect of coupling to an
  external environment.  We use this to argue, both analytically and
  numerically, that the DTC in disordered one-dimensional systems is
  destroyed at long times by any such natural coupling. This holds
  true even in the case where the coupling is such that the system is
  prevented from heating up by an external thermal bath.
\end{abstract}

\maketitle

\newcommand{\allcoords}{\substack{ \omega_{\alpha},\sigma_{\alpha}\\
    \omega_{\beta},\sigma_{\beta}}}

\newcommand{\fourcoords}{\substack{ \omega,\sigma_{\alpha}\\
    \omega,\sigma_{\beta} } }

\newcommand{\threecoords}{
  \substack{\omega\\
    \sigma_{\alpha},\sigma_{\beta} } }

\newcommand{\sign}{\mathrm{sgn}}

\section{Introduction}
The field of non-equilibrium quantum many-body dynamics has seen increasing
interest and rapid progress in recent years.  Periodically-driven, or
Floquet, many-body systems have been among the most rapidly
progressing topics because of the promise of non-trivial
long-time behaviour and the resulting recent experimental
activity.\cite{Bordia2016,Zhang2016a,Choi2016}

The main obstacle to observing interesting physics at long times in
Floquet matter is that a general ergodic system simply heats up
maximally under driving, as its entropy increases due to the
non-adiabatic nature of the
perturbation.\cite{DAlessio:2014fg,Lazarides:2014ie,Ponte:2015hm} This
process may be frustrated either by tuning to certain integrable
points for which the system does not heat up and its long-time
behaviour is described by a so-called ``periodic Gibbs ensemble''
(PGE),\cite{Lazarides:2014cl} or more robustly by introducing disorder
which leads to a many-body localised (MBL) phase in static
systems.\cite{Basko:2006hh,Oganesyan:2007ex,Pal:2010gr} Driving an MBL
system leads to finite-energy-density long-time
states\cite{Lazarides:2015jd,Ponte:2015dc,Bordia2016} which may
display non-equilibrium phases based on the notion of eigenstate
order.\cite{Huse:2013bw,Pekker:2014bj}

One particularly intriguing example is the $\pi-$spin
glass,\cite{Khemani2016} since also known as discrete time crystal
(DTC) \cite{Else:2016ue,VonKeyserlingk2016a,Yao2017}. The
$d+1$-dimensional spatiotemporal order characterising DTCs manifests
itself as a subharmonic response of the system to the driving.  Like
Bragg peaks signalling an increased magnetic unit cell compared to the
structural one upon the onset of antiferromagnetic N\'eel order, this
is encoded in the temporal Fourier transform.

This prediction has immediately sparked activity aimed at the
experimental confirmation of this latest addition to quantum
statistical mechanics.\cite{Zhang2016a,Choi2016} The experiments found
a temporally decaying order parameter. This immediately poses the
question of the stability of the spatiotemporal order in realistic
environmental conditions.  In addition, in this setting there is the
obvious question about the role of decoherence -- a concept of
perennial interest in quantum physics, playing a role in aspects as
fundamental as the measurement process, and applications as important
as quantum
computing.\cite{Hone:2009hs,Vorberg2013,Dehghani:2014jm,Seetharam:2015wj,Shirai2016}

We address these issues by studying the evolution of the density
matrix via a Lindblad equation describing coupling to a Markovian
environment. We find that at least in the one dimensional disordered
systems recently investigated, the spatiotemporal DTC order is
destroyed by any realistic model of environmental effects. We
investigate two generic models of environmental decoherence as well as
a model of an external thermal bath, finding that while the latter
still destroys the DTC it nevertheless does not lead to an
infinite-temperature state.

In detail, our results are the following. First, we show that DTCs can
be described, in the long time limit, by a form of the density matrix
that we call block-diagonal ensemble (BDE), which has a block-diagonal
form in a basis we define related to the Floquet basis and which is
manifestly $2T$ periodic. For any local operator to be sensitive to
this periodicity, disorder must be present and the initial states must
themselves break the $\mathbb{Z}_2$ symmetry of the drive. We then
discuss which properties lead to a Lindblad-type equation which has
this form as a steady state. Because DTC is a property of the Floquet
states, not detectable by spatially and temporally local measurements,
we are led to conclude that Lindblad operators preserving DTC are most
naturally expressed in the Floquet basis and have the property either
of not coupling different Floquet states to each other at all, or of
coupling Floquet states only to others selected based on the local
operator that displays oscillations. This leads to the conclusion that
coupling to physical environments typically destroys DTC, since in
general the environment will not have this property. Thus while it is
possible to write down Lindblad operators preserving the DTC, these do
not appear to correspond to any natural physical processes.

The remainder of this paper is organised as follows. We first
review DTCs in isolated Floquet systems in order to define the
problem and fix notation. We then set up our analysis for the presence
of dephasing non-unitary dynamics. We present our results on the
timescales over which the DTCs persist as well as our general
conclusions on the stability of the DTC in open one dimensional
systems, based on an analysis of different types of coupling models to
the environment.  We conclude with an outlook.

\section{Discrete time crystals: Unitary evolution}
\label{sec:basic_idea_of_time_crystals}

We set up the problem by taking the first and arguably simplest model
of a DTC as our object of study. For a brief review see Ref.~\onlinecite{Moessner}.

\subsection{Model and phenomenology in the $\pi$-SG phase}
\label{sub:setup}

The Floquet dynamics is provided by a binary drive, in which the period $T$ is
subdivided into two parts during each of which a time-independent Hamiltonian
generates unitary evolution:
\begin{equation}
  H(t)=
  \begin{cases}
    H_{z} & \quad\text{if }0\le t<T_{z}\\
    H_{x} & \quad\text{if }T_{x}\le t<T=T_{z}+T_{x}
  \end{cases}\label{eq:hising}
\end{equation}
 The model describing a chain of 
spins-1/2 represented by Pauli matrices $\sigma_i^\gamma$
is\cite{Khemani2016} 
\begin{equation}
  \begin{aligned}
    H_{z} & =\frac{1}{2}\sum_{i=1}^{L}h_{i}\sigma_{i}^{z}+\sum_{i=1}^{L-1}J_{z}\sigma_{i}^{z}\sigma_{i+1}^{z},\\
    H_{x} &
    =\sum_{i=1}^{L-1}J_{i}\sigma_{i}^{x}\sigma_{i+1}^{x}+J_{z}\sigma_{i}^{z}\sigma_{i+1}^{z}
    \label{eq:Hx-Hz}
  \end{aligned}
\end{equation}
Here $J_{i}=1+\delta J_{i}$ with $\delta J_{i}$ randomly drawn from a
uniform distribution between $-\delta J/2$ and $\delta J/2$, while
$h_{i}=\overline{h}+h_{i}$ with $h_{i}$ randomly drawn form a uniform
distribution between $-w/2$ and $w/2$. 

As a technical aside, we note that in the numerics that follows we set
$J_z=0$ unless otherwise indicated as it has been
shown\cite{Khemani2016,VonKeyserlingk2016a} that the DTC phase is
stable to the introduction of interactions. However the numerically
accessible range of $J_z$ can be quite small especially as our exact
approach only allows for small system sizes for which a flip of
a single spin's $z$ component with concomitant energy change $J_z$ can
already amount to considerable energy ``density.''

The unitary operator
\begin{equation}
  U:=U(T,0)=\exp\left(-iH_{x}T_{x}\right)\exp\left(-iH_{z}T_{z}\right)
  \label{ugeneral}
\end{equation}
propagates the system over one period. Its eigenvectors $\ket{\omega_\alpha}$,
$U\ket{\omega_{\alpha}}=\exp\left(-i\omega_{\alpha}T\right)\ket{\omega_{\alpha}},$
are the Floquet states which for stroboscopic dynamics play a role
analogous to that played by energy eigenstates for static systems.
The eigenvalues are of the form $\exp\left(-i\omega_{\alpha}T\right)$
with $T=T_{x}+T_{z}$ the period and the quasienergies $\omega_{\alpha}$
real. For later convenience let us also define
\begin{equation}
  \ket{\omega_{\alpha};t}=U(t,0)\ket{\omega_{\alpha}}
  \label{eq:floquet-state-arbitrary-time}
\end{equation}
 which is useful in describing in-period dynamics.

\begin{figure}[t]
  \includegraphics[width=1\columnwidth]{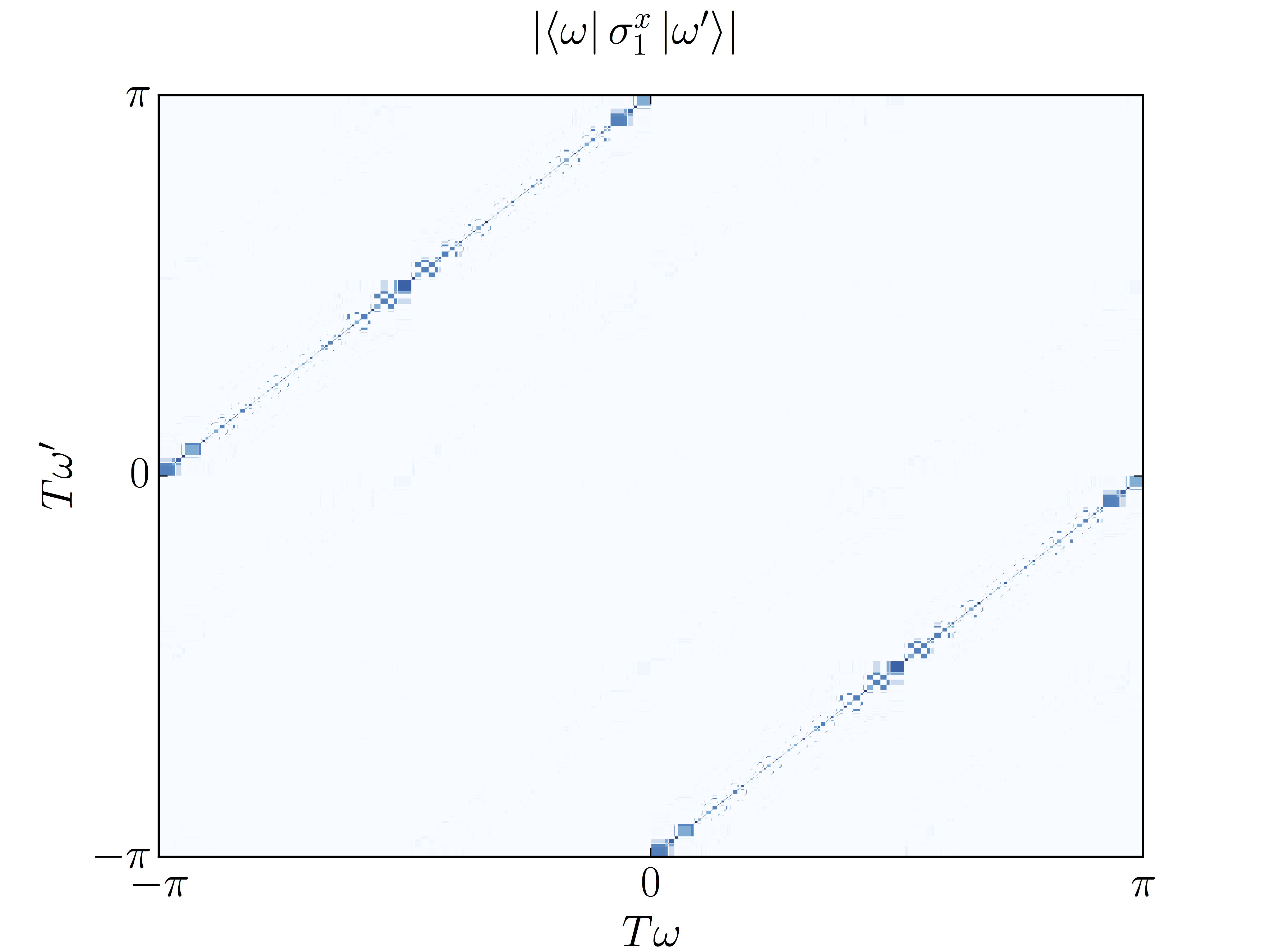}
  \caption{Magnitude of matrix element
    $\left<\omega\right|\sigma^x_1\left|\omega'\right>$ for a single
    disorder realisation and a system size $L=10$ of the system of
    Eq.~(\ref{eq:hising}) for bond- and site-disorder amplitudes
    $\delta J=w=0.1J$ versus the quasienergies of the Floquet states
    between which it is taken. Darker colour indicates larger magnitude. The weight
    is mostly concentrated at elements between Floquet states
    differing by quasienergy $\pi/T$, implying that the dynamics
    of the operator $\sigma^x_1$ will display a strong oscillation at
    frequency $\pi/T$ or period $2T$ as in
    Eq.~(\ref{eq:oscillatory-sx}).
    \label{fig:matrix-els-sigma-x}}
\end{figure}

The model of Eq.~\eqref{eq:hising} supports a number of phases; in
this paper we shall be exclusively concerned with its time-crystalline
$\pi$-SG phase, in which the eigenstates of $U$ have the following
salient
properties:\cite{Thakurathi:2013dt,Benito:2014bd,Khemani2016,VonKeyserlingk2016a}
 \begin{enumerate}
 \item Are eigenstates of the parity operator
   \begin{equation}
     P=\prod_{j}\sigma_{j}^{z}.
     \label{eq:parity}
   \end{equation}
 \item Come in pairs of opposite parity, separated by quasienergy
   $\pi/T$ (up to corrections exponentially small in the system size)
   and forming a \emph{doublet} for each $\omega$. They can be
   labelled as $\ket{\omega_{\alpha},\pm}$ with
   \begin{equation}
     \label{eq:omega-plus-minus}
     U\ket{\omega_{\alpha},\pm}=\pm\exp(-i\omega_\alpha T)\ket{\omega_{\alpha},\pm}.
   \end{equation}
   The doublet can also be thought of as a \emph{pseudospin}-1/2 for each
   $\omega$ block.
 \item Display spatial long-range order: the expectation value of the operator
   $\sigma^x_i\sigma^x_j$ is finite in all eigenstates for arbitrarily
   large $|i-j|$ and is the same for both members of each
   doublet.
        \begin{equation}
       \label{eq:LRO}
       \bra{\omega,+}\sigma^x_i\sigma^x_j\ket{\omega,+}
       =
       \bra{\omega,-}\sigma^x_i\sigma^x_j\ket{\omega,-}
       \neq 0
     \end{equation}
     It is however different for different $\omega$, and does not
     depend smoothly on $\omega$.
   \item The matrix elements of the operator $\sigma^x_j$
     are finite between the two members of a doublet as shown in
     Fig.~\ref{fig:matrix-els-sigma-x}. 
 \end{enumerate}

 The last point above indicates that the dynamics of this operator
 will exhibit subharmonic
 oscillations. To bring this out more clearly, let us rotate the
 doublet basis $\ket{\pm}$ to a broken symmetry basis,\footnote{Note
   that $\uparrow/\downarrow$ do not refer to physical spins; they
   simply label the two many-body states.}
\begin{equation}
  \ket{\uparrow/\downarrow}=\left(\ket{+}\pm\ket{-}\right)/\sqrt{2}.
  \label{eq:updown-basis}
\end{equation}
These two basis states are eigenstates neither of the parity operator
$P$ nor of the unitary operator $U$ and have the following
properties:
\begin{enumerate}
\item The expectation value of $\sigma^x_j$ is generically finite in
  each of these.
\item From Eq.~\eqref{eq:omega-plus-minus} it follows that the action
  of $U$ is to flip the spin and produce a phase:
  \begin{equation}
    \label{eq:u-on-up-down}
    U\ket{\omega,\uparrow/\downarrow}
    = \exp\left(-i\omega T \right)\ket{\omega,\downarrow/\uparrow}
  \end{equation}
\item From Eq.~\eqref{eq:updown-basis} it follows that the action of
  the parity operator is to flip the spin:
    \begin{equation}
    \label{eq:p-on-up-down}
    P\ket{\omega,\uparrow/\downarrow}
    = \ket{\omega,\downarrow/\uparrow}
  \end{equation}
\item From the two last points one concludes that the broken symmetry
  states break spatiotemporal symmetry, with a subharmonic response of
  period $2T$: if
  $\ket{\psi_{0}}=\ket{\omega_{\alpha},\uparrow}$ and
  $\ket{\psi_{m}}=U^{m}\ket{\psi_{0}}$ for integer $m$ then
\begin{equation}
  \begin{split}
    \bra{\psi_{2n}}\sigma_{j}^{x}\ket{\psi_{2n}}
    & =\bra{\psi_{0}}\sigma_{j}^{x}\ket{\psi_{0}}\\
    \bra{\psi_{2n+1}}\sigma_{j}^{x}\ket{\psi_{2n+1}} &
    =-\bra{\psi_{0}}\sigma_{j}^{x}\ket{\psi_{0}}
  \end{split}
  \label{eq:oscillatory-sx}
\end{equation}
\end{enumerate}

\section{The long-time limit and the block-diagonal ensemble}
\label{sec:Description-in-terms-of-density-matrices}

In this section we  use the above properties to show that the
long-time state of the system is well described by a density matrix of
a special form that we call the ``block-diagonal ensemble''
(BDE). This form follows naturally from the doublet structure and the
form of the matrix elements of operators breaking the $\mathbb{Z}_2$
symmetry (Fig.~\ref{fig:matrix-els-sigma-x}) and generalises the
diagonal ensemble occuring in systems with no special spectral
structure.\cite{Russomanno:2012bf,Lazarides:2014cl,Moessner}

Let us now consider the form of the BDE matrix in detail. An initial
density operator with matrix elements
$\rho_{\alpha\beta}=\bra{\omega_\alpha}\rho\ket{\omega_\beta}$ after
$n$ periods becomes
$\rho(nT)=\sum_{\alpha,\beta}\rho_{\alpha\beta}\exp\left(-i\left(\omega_{\alpha}-\omega_{\beta}\right)nT\right)\ket{\omega_\alpha}\bra{\omega_\beta}$.
For a generic (ergodic, non-MBL, non-spectrally-paired) system, a) the
eigenvalues $\omega_{\alpha}$ are continuously distributed with no
special spectral structure apart from repelling each other due to the
ergodic nature of the system and b) local operators have matrix
elements that are maximal near the diagonal
$\omega_\alpha=\omega_\beta$, as for static
systems.\cite{Beugeling2015} This leads to such operators having
synchronised states which are described by the so-called diagonal
ensemble\cite{Rigol:2008bf} in the floquet basis, in which terms with
$\alpha\neq\beta$ do not contribute at long times. The result is that
the long-time steady-state becomes independent of the period $n$, thus
synchronised with the
driving.\cite{Russomanno:2012bf,DAlessio:2014fg,Lazarides:2014cl,Ponte:2015hm}

In the $\pi-$SG case, by contrast, a) the eigenvalues are continuously
distributed \emph{except} for the doublet pairing structure, and b)
there exist operators the matrix elements of which are appreciable
between pairs of states separated  by quasienergy
$\omega/2=\pi/T$.  If the operators of interest possess the latter
property, terms off-diagonal in the Floquet basis and differing by
quasienergy $\pi/T$ are the only ones that survive leading to what we
call the ``Block Diagonal Ensemble'' (BDE). Density matrices of
this form describe a steady state which is periodic with period twice
that of the driving (rather than with the same period as the driving).

\begin{widetext}
  Concretely, in terms of the broken-symmetry states a general
  initial density matrix may be written as
\begin{equation}
  \rho_{0}=\sum_{\allcoords}\rho_{\allcoords}\ket{\omega_{\alpha},\sigma_{\alpha}}
  \bra{\omega_{\beta},\sigma_{\beta}}\label{eq:density-matrix-general}
\end{equation}
which may be visually represented as 
\begin{equation}
  \rho_{0}=\begin{array}{c|c|c|c|c}
             \ddots &  &  & \\
             \hline  & \rhospinor{\alpha-1}{\alpha-1} & \rhospinor{\alpha-1}{\alpha} &\rhospinor{\alpha-1}{\alpha+1}\\
             \hline  & \rhospinor{\alpha}{\alpha-1} &\rhospinor{\alpha}{\alpha} &  \rhospinor{\alpha}{\alpha+1}\\
             \hline  & \rhospinor{\alpha+1}{\alpha-1} & \rhospinor{\alpha+1}{\alpha} & \rhospinor{\alpha+1}{\alpha+1}\\
             \hline  &  &  &  & \ddots
           \end{array}\label{eq:full-density-matrix-pict}
\end{equation}
with the matrices
\begin{equation}
  \rhospinor{\alpha}{\beta}
  =
  \left(\begin{array}{cc}
          \rho_{\omega_{\alpha,\omega_{\beta}};\uparrow,\uparrow} & \rho_{\omega_{\alpha,}\omega_{\beta};\uparrow,\downarrow}\\
          \rho_{\omega_{\alpha,}\omega_{\beta};\downarrow,\uparrow} &
                                                                      \rho_{\omega_{\alpha,}\omega_{\beta};\downarrow,\downarrow}
        \end{array}\right).
      \label{eq:original-spinor}
    \end{equation}
Time evolved (see Eq.~\eqref{eq:omega-plus-minus}) this becomes
$\rho_{n}=U^{n}\rho(U^{\dagger})^{n}$,
  \[
    \rho_{n}=\sum_{\allcoords}\rho_{\allcoords}\mathrm{e}^{\left(-i(\omega_{\alpha}-\omega_{\beta})nT\right)}\ket{\omega_{\alpha},(-1)^{n}\sigma_{\alpha}}\bra{\omega_{\beta},(-1)^{n}\sigma_{\beta}}.
  \]
  The expectation value of $\sigma_{j}^{x}$
  after $n$ periods is
\[
  \mathrm{tr}\left(\rho_{n}\sigma_{j}^{x}\right)=\left(-1\right)^{n}\sum_{\allcoords}\rho_{\allcoords}\bra{\omega_{\alpha},\sigma_{\alpha}}\sigma_{j}^{x}\ket{\omega_{\beta},\sigma_{\beta}}\exp\left(-i\left(\omega_{\alpha}-\omega_{\beta}\right)nT\right).
\]
That is, time evolution affects parts of this sum corresponding to the
diagonal blocks ($\alpha=\beta$) of the representation in
Eq.~\ref{eq:full-density-matrix-pict} only via a period-$2T$ flipping
of the pseudospin, while all the other blocks acquire an additional
time-dependent phase. 

These phases are not correlated with each other,
nor are they commensurate with the driving. Noting that \emph{all}
diagonal blocks give a $2T$-periodic contribution to the sum, while
off-diagonal blocks also contribute a ``random'' frequency
(incommensurate with that of the driving and the other blocks), we
conclude (as earlier) that the off-diagonal blocks may be set to zero in the
long-time limit for the same reason the diagonal ensemble holds in the
absence of the doublet structure. This then leads to the BDE,
\begin{equation}
  \rho^{BDE}=\sum_{\threecoords}\rho_{\omega;\sigma_{\alpha}\sigma_{\beta}}\ket{\omega,\sigma_{\alpha}}\bra{\omega,\sigma_{\beta}}
  \label{eq:bde-symbols}
\end{equation}
visually represented as
\begin{equation}
  \rho^{\mathrm{BDE}}=\begin{array}{c|c|c|c|c}
                        \ddots &  &  & \\
                        \hline  & \tilde{\mathbf{\rho}}_{\alpha-1,\alpha-1} & 0 & 0\\
                        \hline  & 0 & \mathbf{\tilde{\rho}}_{\alpha,\alpha} & 0\\
                        \hline  & 0 & 0 & \mathbf{\tilde{\rho}}_{\alpha+1,\alpha+1}\\
                        \hline  &  &  &  & \ddots
                      \end{array}\label{eq:bde-form}
\end{equation}
\end{widetext}
The action of $U$ on this is only to flip $\uparrow/\downarrow$ to
$\downarrow/\uparrow$, that is,
\begin{equation}
  U \rho^{BDE}
  U^\dagger=\sum_{\threecoords}\rho_{\omega;\sigma_{\alpha}\sigma_{\beta}}
  \ket{\omega,-\sigma_{\alpha}}\bra{\omega,-\sigma_{\beta}}
  \label{eq:bde-flipped-symbols}
\end{equation}
Hence, provided that
$\rho_{\omega_{\alpha},\omega_{\alpha};\downarrow,\uparrow}\neq\rho_{\omega_{\alpha},\omega_{\alpha,};\uparrow,\downarrow}$
or
$\rho_{\omega_{\alpha},\omega_{\alpha};\downarrow,\downarrow}\neq\rho_{\omega_{\alpha},\omega_{\alpha};\uparrow,\uparrow}$
(or both), a density matrix of the form Eq.~\ref{eq:bde-form} has
period $2T$, since
$U^{2n}\rho^{\mathrm{BDE}}(U^{\dagger})^{2n}=\rho^{\mathrm{BDE}}$ for
any integer $n$ while
$U^{2n+1}\rho^{\mathrm{BDE}}(U^{\dagger})^{2n+1}\neq
\rho^{\mathrm{BDE}}$.  In particular, we note that the expectation
value $\sigma_{j}^{x}$ in the BDE is
\[
\mathrm{tr}(\rho_{n}^{BDE}\sigma_{j}^{x})=\left(-1\right)^{n}\mathrm{tr}(\rho_{0}^{BDE}\sigma_{j}^{x})
\]
from which it follows that \emph{the spin oscillates with period $2T$ in
the BDE}.

For finite system sizes $L$, the two partner states are not separated
by quasienergy exactly $\pi/T$ but rather by $\pi/T+\epsilon$ with
$\epsilon$ exponentially small in the system size, and different for
each pair of states. After a time $\sim\exp(L)$, which diverges in the
thermodynamic limit, the oscillations are therefore washed out and the
DTC disappears.\cite{VonKeyserlingk2016a}

\section{Lindblad operators and the BDE form}
\label{subsec:Lindblad-operators-and-BDE-form}

The most general equation for the possibly non-unitary evolution of a quantum density
matrix is given by the Lindblad equation (see Appendix~\ref{sec:the_lindblad_equation})
\begin{equation}
  \partial_{t}\rho=\mathcal{L}\rho
  \label{eq:Lindblad}
\end{equation}
with the Lindblad operator 
\[
  \mathcal{L}\rho=-i\left[H,\rho\right]+\sum_{a}\left(L_{a}\rho
    L_{a}^{\dagger}-\frac{1}{2}\left[L_{a}^{\dagger}L_{a},\rho\right]_{+}\right)
\]
where $\left[\cdot,\cdot\right]_{+}$ is the anticommutator and the
$L_a$ arbitrary operators encoding the non-unitary part of the dynamics. In
what follows, we  use this form with the time-dependent
Hamiltonian of Eq.~(\ref{eq:hising}) and choices of Lindblad
operators $L_a$ appropriate to different types of environmental
couplings.

We begin by discussing which features the operators $L_a$ must display
in order to not destroy the BDE structure responsible for the
subharmonic oscillations. As the DTC is a phenomenon that cannot be
detected by local measurements at a single point in time, the Lindblad
operators that preserve DTC are naturally written down in terms of
Floquet operators. We will identify properties required for the DTC to
survive and argue that these are unlikely to appear in real
systems.

The Lindblad equations we consider are of the form
$\partial_{t}\rho=\mathcal{L}\left(t\right)\rho$ with 
\begin{equation}
  \label{eq:lindblad-floquet}
    \mathcal{L}(t)\rho
    =-i\left[H(t),\rho\right] + \mathcal{D}(\rho)
\end{equation}
with 
\begin{equation*}
  \mathcal{D}(\rho)
  =
  \sum_{\omega,\omega'}\left(L_{\omega,\omega'}\rho
      L_{\omega,\omega'}^{\dagger}-\frac{1}{2}\left[L_{\omega,\omega'}^{\dagger}L_{\omega,\omega'},\rho\right]_{+}\right).
\end{equation*}
We take the Lindblad operators to be of the general form
\begin{equation}
  \label{eq:Lomomprime}
  \begin{split}
    L_{\omega,\omega'}(t)&
    =\sqrt{\gamma_{\omega,\omega'}}\sum_{\sigma}
    \ket{\omega,\sigma;t}\bra{\omega',\sigma;t}\\
    &=\sqrt{\gamma_{\omega,\omega'}}\ket{\omega;t}\bra{\omega';t}\otimes\mathbb{I}.
  \end{split}
\end{equation}
Explicitly, defining
\newcommand{\rhohat}{\mathbf{\hat{\rho}}}
\begin{equation}
  \label{eq:rhohat}
  \rhohat_{\omega_\alpha,\omega\beta}=\sum_{\sigma_\alpha,\sigma_\beta}\rho_{\allcoords}\ket{\sigma_\alpha}\bra{\sigma_\beta},
\end{equation}
we have
\begin{equation}
  \label{eq:block-mixing-lindblad-matrix}
  \partial_t \rhohat_{\omega_\alpha,\omega_\beta}
  =
  -i\left[H(t),\rhohat_{\omega_\alpha,\omega_\beta}\right]
  + \mathcal{D}_{\omega_\alpha,\omega_\beta} 
\end{equation}
with
\begin{equation}
  \label{eq:rhs-master-equation}
  \begin{split}
  \mathcal{D}_{\omega_\alpha,\omega_\beta}
  &=
  \delta_{\alpha,\beta}\sum_{\delta}
  \gamma_{\omega_\alpha,\omega_\delta}  \rhohat_{\omega_\delta,\omega_\delta}\\
  &\quad - \frac{1}{2} \rhohat_{\omega_\alpha,\omega_\beta}
  \sum_\delta
  \left(\gamma_{\omega_\delta,\omega_\alpha} +
    \gamma_{\omega_\delta,\omega_\beta}\right)
\end{split}
\end{equation}
which manifestly has steady states of the BDE form of
Eq.~(\ref{eq:bde-symbols}): Intra-block dynamics is only generated by
the unitary term, which as discussed in
Sec.~\ref{sec:Description-in-terms-of-density-matrices} results in the
subharmonic features, while inter-block dynamics are only generated by
the $\mathcal{D}$ term. The time evolution due to the latter couples
the diagonal blocks ($\alpha=\beta$) to each other and only to each
other, while each off-diagonal block is not coupled
($\alpha\neq\beta$) to any other blocks so that each simply decays
exponentially with time.

\subsection{No mixing between blocks}
\label{sec:no-mixing-blocks}

We begin with a DTC-preserving choice for the Lindblad operators
of~\eqref{eq:Lomomprime}, namely,
$\gamma_{\omega,\omega'}=\gamma_\omega\delta_{\omega,\omega'}$ so that
$L_{\omega,\omega'}(t)=\sqrt{\gamma_\omega}
\delta_{\omega,\omega'}\ket{\omega;t}\bra{\omega;t}\otimes\mathbb{I}$
is a projector onto a doublet.  For this projector form
$\mathcal{D}_{\omega_\alpha,\omega_\alpha}=0$ while for
$\alpha\neq\beta$ we have
$\mathcal{D}_{\omega_\alpha,\omega\beta}=- \frac{1}{2}
\rhohat_{\omega_\alpha,\omega_\beta} \left(\gamma_{\omega_\alpha} +
  \gamma_{\omega_\beta}\right)$. Therefore the entire time dependence
is exponential damping of all off-diagonal blocks (with rates
determined by the various decoherence channels) while the diagonal
ones are completely unaffected by the evolution. Consequently, any
initial density matrix with pseudospin imbalance (thus an initial
finite expectation value of $\sigma^x_j$) will result in a BDE
long-time density matrix displaying subharmonic oscillations.

While we have succeeded in constructing Lindblad operators leading to
the BDE form and subharmonic oscillations, we still need to discuss
how realistic such a choice of operators is. To answer this, it is
useful to think of the process described by Eq.~(\ref{eq:Lindblad}) as
the system evolving unitarily, with the unitary evolution interrupted
at random and with a typical rate set by the $\gamma$, by measurements
described by the operators
$L_{\omega,\omega'}$.\cite{Dalibard:1992dv,Carmichael:1993el} For the
environment to have the effect described here it must effectively
measure the occupancies of doublets. This is not a natural operation
for two reasons: Firstly, a measurement of a projector onto a doublet
is impossible to achieve with local operations as it would require a
simultaneous measurement at all points in space; secondly, the
projector is time-dependent, in a way dictated by the unitary time
evolution of the Floquet system itself.  Such an external environment
therefore seems very fine-tuned and unlikely to appear naturally, even
if it is not physically forbidden.

\subsection{Mixing between blocks}
\label{sec:mixing-blocks}

One may ask whether it is possible to remove the restriction of no
mixing between the blocks and have DTC still survive. Here we show
that the answer is yes for certain fine-tuned choices of Lindblad
operators; for general forms of the
$\gamma_{\omega_\alpha,\omega_\beta}$ the DTC is destroyed.

To lighten the notation we use the fact that the off-diagonal blocks
evolving according to Eq.~\eqref{eq:block-mixing-lindblad-matrix}
simply decay exponentially, so the long-time BDE state may be obtained
by solving the equation
\begin{equation*}
  \partial_t \rhohat_{\omega_\alpha} =
  -i\left[H(t),\rhohat_{\omega_\alpha}\right] + 
  \sum_\epsilon\left(
    \rhohat_{\omega_\epsilon}\gamma_{\omega_\alpha,\omega_\epsilon} -
    \rhohat_{\omega_\alpha}\gamma_{\omega_\epsilon,\omega_\alpha}
  \right)
\end{equation*}
with $\rhohat_{\omega_\alpha}=\rhohat_{\omega_\alpha,\omega_\alpha}$
(see Eq.~\eqref{eq:rhohat}. As before, the unitary (first) term only
produces intra-block dynamics by flipping the pseudospin periodically;
the interblock dynamics is described by a master equation for the
$\rhohat_{\omega}$. The steady state is determined by setting the sum
on the right hand side to zero and is automatically of the BDE form.
For general $\gamma$ that do not separate the state space into
disjoint sets, such master equations generally result in a steady
state in which all the blocks have a finite occupancy, with the
occupancies determined by the details of the $\gamma$. This will in
general result in vanishing expectation values for $\sigma^x_j$ due to
the fact that its expectation value varies randomly between blocks and
is not correlated with the pseudospin; this follows from the
discussion around Eq.~\eqref{eq:LRO} in section~\ref{sub:setup}. We
now discuss when this response might not vanish, finding that the
conditions again correspond to environments with very unnatural
properties.

First, there will be a subharmonic response if the steady state
corresponds to a single occupied block such that there is pseudospin
imbalance. A single-block steady state will occur if the $\gamma$ are
such that there is a state into which there are transitions but out of
which there are no transitions. This would require the effect of the
environment to favour one particular floquet state over all others,
corresponding again to an environment performing fine-tuned,
time-dependent operations everywhere in space simultaneously.

Second, a subharmonic oscillation of the operator $\sigma^x_j$ for a
given $j$ will also result if the $\gamma$ only connect blocks for
which, for the same pseudospin orientation for each block,
$\sigma^x_j$ has the same sign. In this case the sign of the spin will
flip in time, while its magnitude may decrease but will not
vanish. However, for a finite fraction, say $\nu$, of the spins to
oscillate, the $\gamma$ must only connect a fraction $2^{-\nu L}$ of
the blocks which vanishes in the thermodynamic limit. Thus the
environment would need to be selective in which states it couples,
which is the same problem as in our earlier attempts to construct
DTC-preserving operators. In addition this scheme suffers from all the
drawbacks mentioned earlier, namely, that the environment would need
to be performing finely-tuned time-dependent and non-local
measurements on the system.

\begin{figure*}
  \includegraphics[width=1\columnwidth]{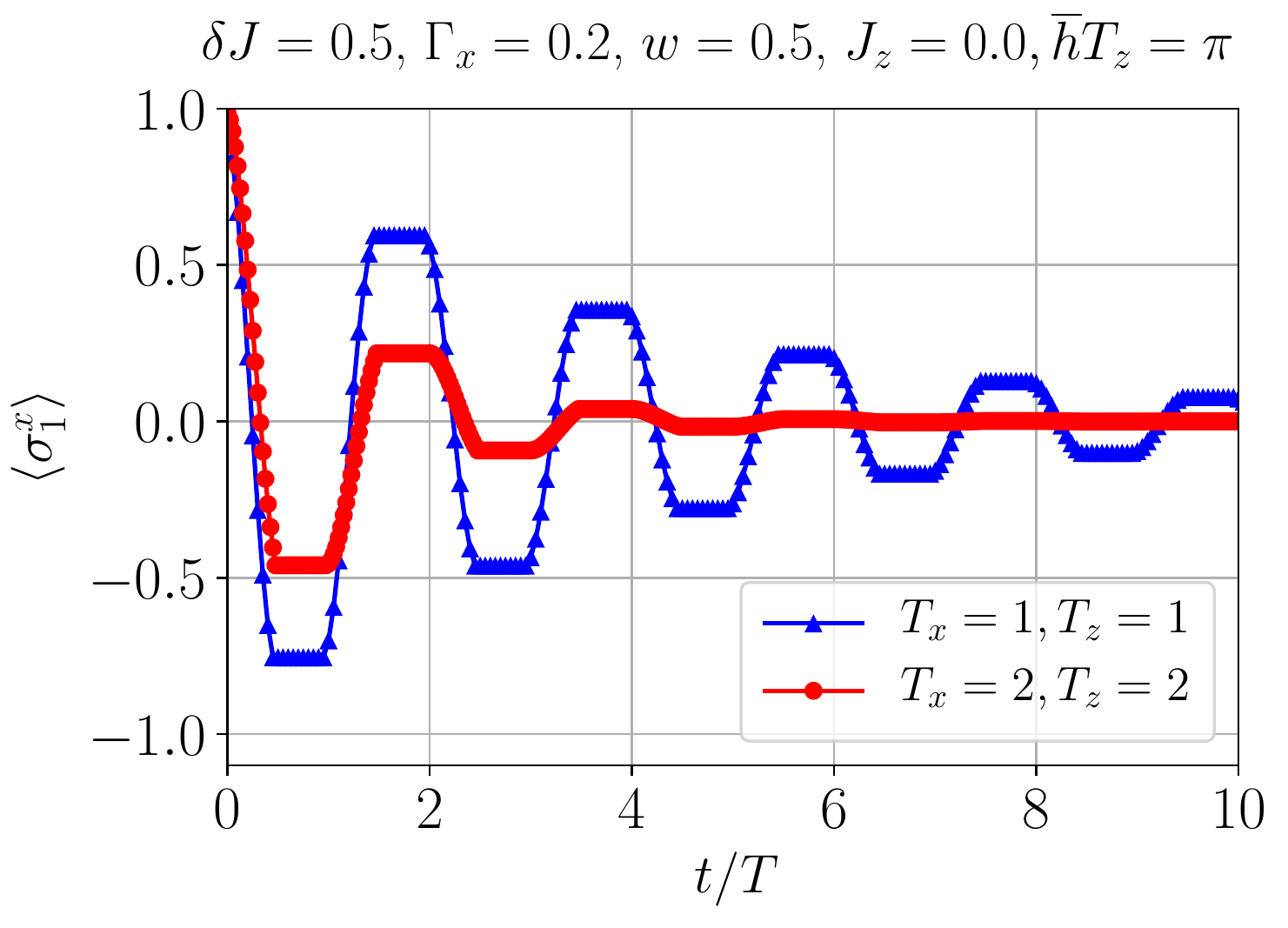}
  \includegraphics[width=1\columnwidth]{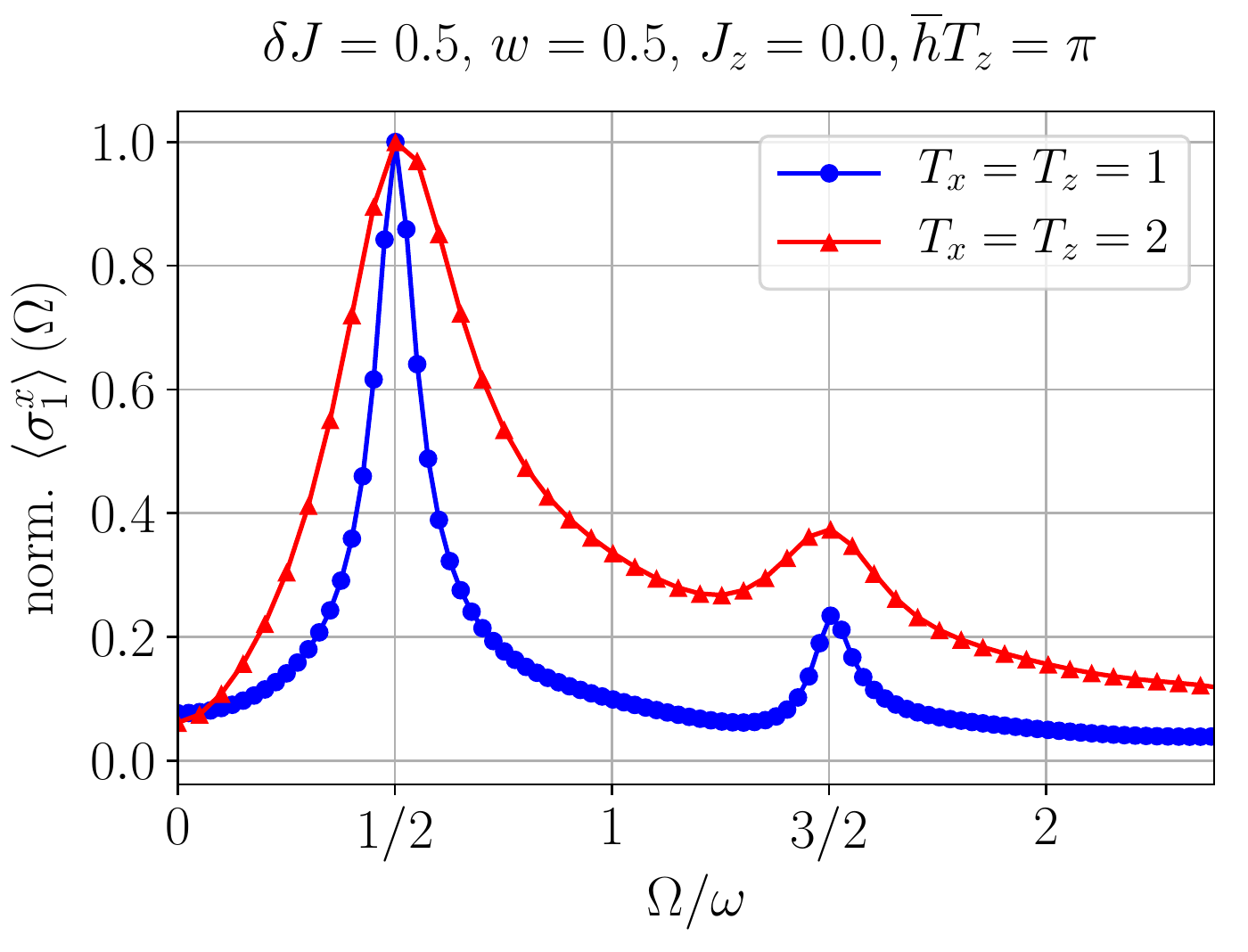}
  \caption{ \textbf{Left}: Real time plot of
    $\left\langle \sigma_{1}^{x}\right\rangle \left(t\right)$ using
    the Lindblad dephasing operators of
    Eq.~\ref{eq:dephasing-lindblad-x}.  The average $h_{i}$, denoted
    by $\bar{h}$ is chosen such that $\bar{h}T_{z}=\pi$, while each
    site has a different, random $h_{i}$. The blue (triangular) and
    red (circular) markers indicate the choice $T_{x}=T_{z}=1$
    ($T_{x}=T_{z}=4$), and the time axis is rescaled by
    $T=T_{x}+T_{z}$ to allow comparison of the two sets. The main
    feature is the longer $T_{z}$ is the faster the decay
    of the spatiotemporal order, as explained in
    Sec.~\ref{sec:Analytical-approach}. The data displayed is for size
    $L=6$. Note that the decay of the DTC occurs during the part of
    the period where the spins are rotating, while there is no decay
    during the part where the plateaus are in contrast to the
    results of Fig.~\ref{fig:Real-time-plot-z-dephasing}
    \textbf{Right}: Fourier transform of the time evolution,
    normalised so that the peaks have the same height to facilitate
    comparison of the width. The important features are that a) both
    cases have a peak at half the driving frequency, b) slower driving
    results in a broader frequency (corresponding to a faster-decaying
    oscillation in real time, as on the left panel).
    \label{fig:Real-time-plot-x-dephasing}}
\end{figure*}

\begin{figure}
  \centering \includegraphics[width=\columnwidth]{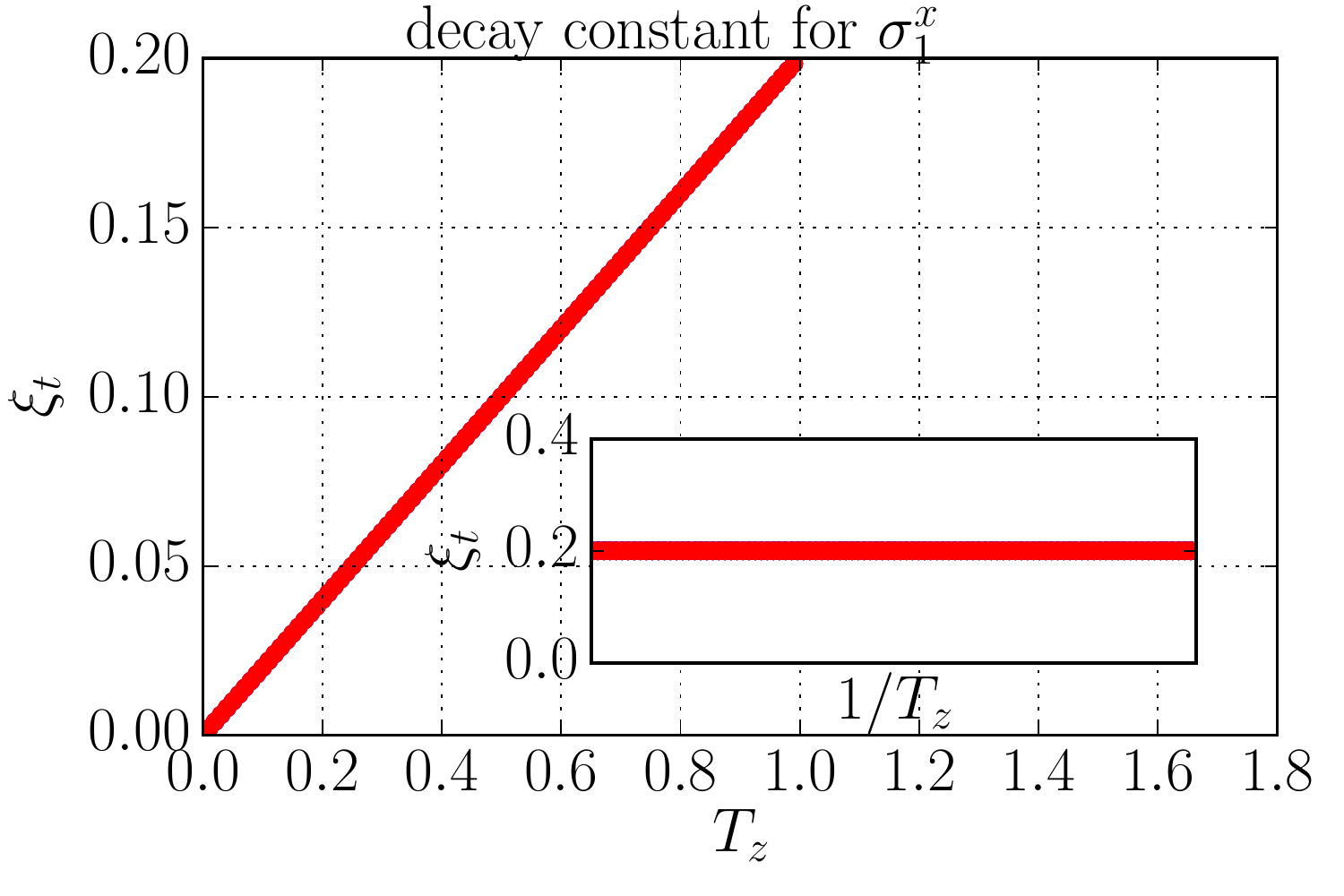}
  \caption{Main panel: Decay rates versus $T_{z}$, the time over
    which the $\pi-$rotation is effected, for sizes $L=4,5$ in blue,
    red respectively (the data points coincide so the blue are not
    visible). Here, $J=T_{x}=1,\delta J=0.2,\Gamma=0.2$ and the data
    has been averaged over 100 (400) disorder realizations for
    $L=4(5)$.  Inset: Same as left, but with $\Gamma T_{z}$ kept
    constant (by varying $\Gamma$). The decay rate is constant,
    indicating that the decay rate is a function of the product
    $\Gamma T_{z}$ only. This is consistent with the results of
    \ref{sec:Analytical-approach}. The data has been averaged over 200
    disorder realisations for both $L=4,5$.}
\label{fig:decay-rate-plots-tau} 
\end{figure}

\section{Physically motivated Lindblad operators}
\label{sec:exp-rel-lind-ops}

We now turn to the direct numerical solution of the Lindblad equation
for three commonly studied and physically realizable types of Lindblad
operators. Two of them are dephasing operators, corresponding to
generic interactions with the environment that destroy quantum
coherence and have no particular energy structure. The third  are
``thermal'' operators, which in the static case  lead to thermal
(Gibbs) long-time states. In all three cases we find that DTC is
destroyed, in agreement with the arguments of
Sec.~\ref{subsec:Lindblad-operators-and-BDE-form}, since these
operators do not have the special structure required to preserve the
BDE form.

In what follows we set $h_{i}$ and $T_{z}$ such that the average
$h_{i}$ is $\overline{h}$ with $\overline{h}T_{z}=\pi$.  In all cases
we take as initial state an equal superposition of the ground and
first excited states of an Ising Hamiltonian in the ferromagnetic
phase, with $J=1$ and $h=0.2$. This ensures that $\sigma_{1}^{x}$ has a
finite expectation value (the site $i=1$ is arbitrarily selected--our
results are independent of this choice).

\begin{figure*}
  \includegraphics[width=1\columnwidth]{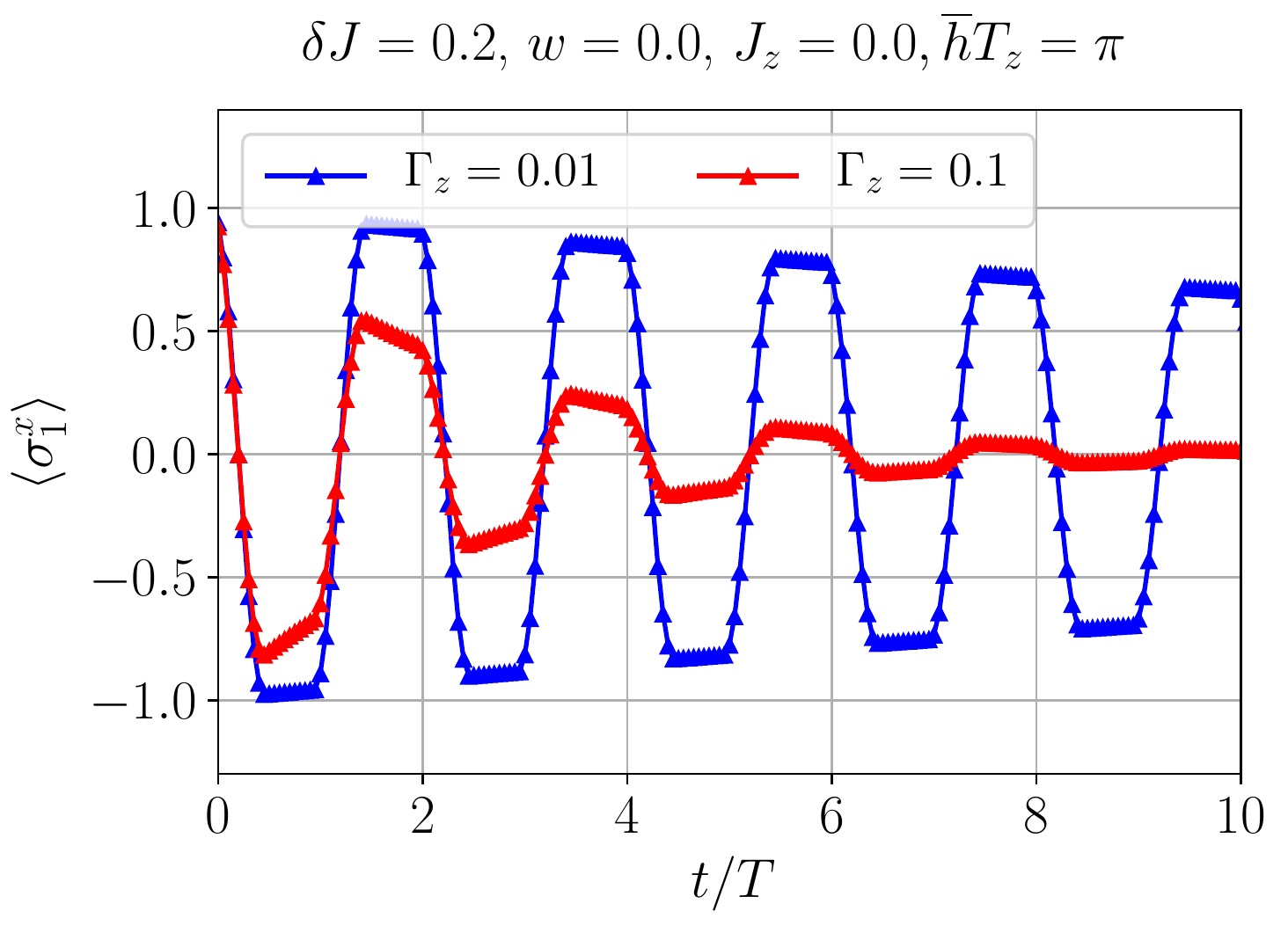}
  \includegraphics[width=1\columnwidth]{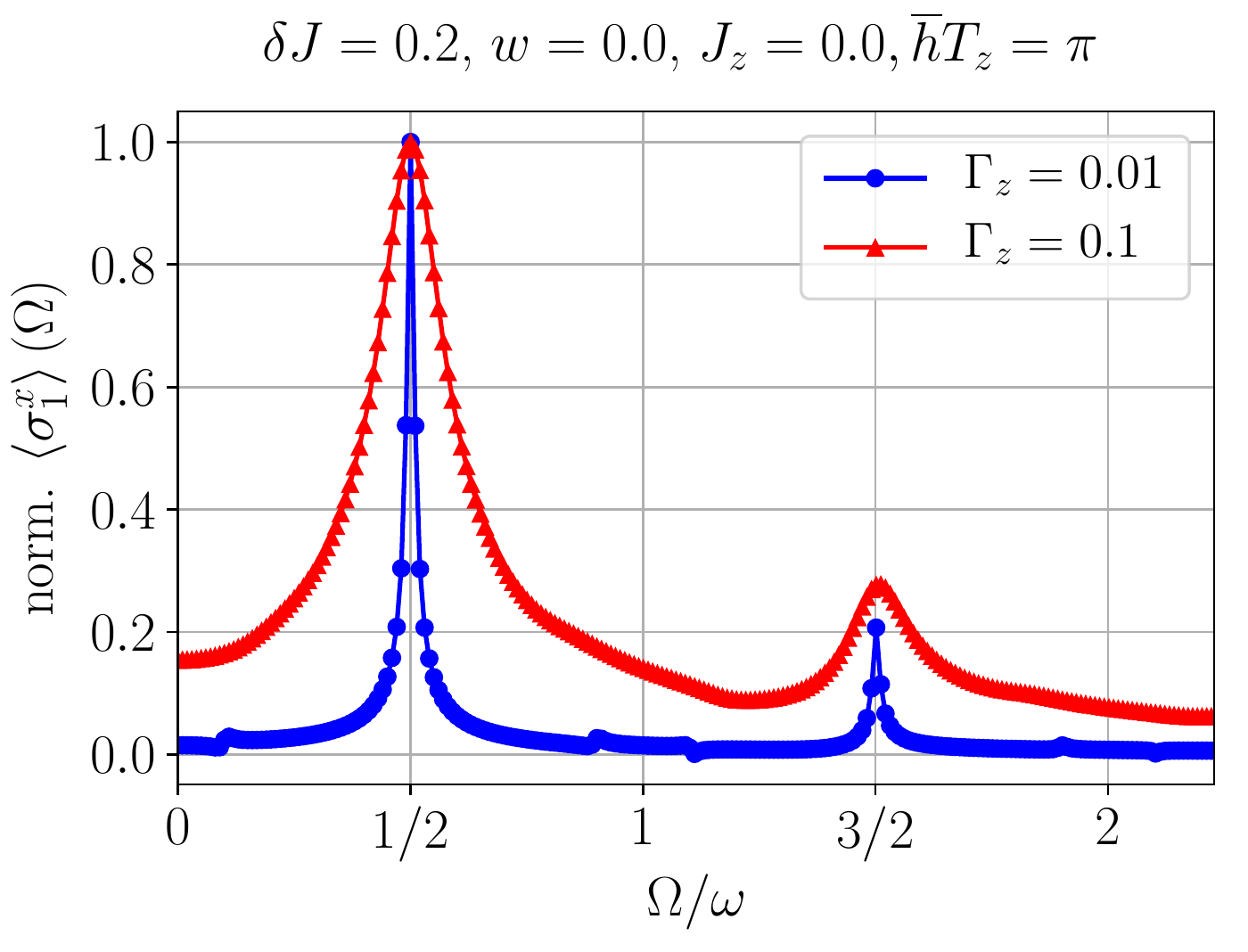}
  \caption{\textbf{Left}:
    Real time plot of
    $\left\langle \sigma_{1}^{x}\right\rangle \left(t\right)$ using
    the Lindblad dephasing operators of
    Eq.~\ref{eq:dephasing-lindblad-z}, for times (see
    Eq.~\ref{eq:hising}) $T_{x}=T_{z}=1$. The blue (triangular) and
    red (circular) markers indicate different dephasing rates. The
    data displayed is for size $L=5$. The main qualitative difference
    from Fig.~\ref{fig:Real-time-plot-x-dephasing} is that the decay
    of the oscillations occurs also during the part of the period
    where the spins are aligned with the $x$ axis (indicated by the
    sloped plateaus in this figure and the flat plateaus in
    Fig.~\ref{fig:Real-time-plot-x-dephasing}).
    \textbf{Right}:
    Fourier transform of the data in the left panel. The frequency
    $\Omega$ is scaled by the driving frequency $\omega$ and the
    vertical axis is scaled so that the highest value of each trace is
    $1$ to make the broadening easier to see. Note that, first, the
    peak is at $\omega/2$ or half the driving frequency, and second,
    the spectrum broadens for stronger dephasing.
    \label{fig:Real-time-plot-z-dephasing}}
\end{figure*}

\subsection{Dephasing}

We start by looking at dephasing operators, relevant to trapped ion
and cold atom experiments as well as to the experiment reported in
Ref.~\onlinecite{Choi2016}. Physically, they model the effect of an
external environment that performs projective measurements of the spin
in some direction. Alternatively, they model environmental effects
without a preferred energy scale.  One example is the non-unitary
dynamics generated by incoherent scattering of the lattice laser light
in cold atom systems.\cite{Pichler:2010cs}

The two cases we study differ in that one preserves the parity of the
initial state while the other does not; however, as we will see, the
DTC is destroyed in both cases.

\subsubsection{Parity-violating dephasing:  $x$ direction}
\label{sec:dephasing-x}

The first type of dephasing operator we consider is
\begin{equation}
  L_{j}=\sqrt{\Gamma}\sigma_{j}^{x}
  \label{eq:dephasing-lindblad-x}
\end{equation}
for $j=1,2,\ldots,N$, aligned along the ferromagnetic direction of
Eq.~\eqref{eq:hising}. Real-time (as opposed to stroboscopic) results
are shown in Fig.~\ref{fig:Real-time-plot-x-dephasing}. The figure
shows $\sigma^{x}$ at an arbitrarily selected position for fixed
values of $\Gamma$ and $\bar{h}T_{z}$ and varying $T_{x}$ and $T_{z}$.

The Lindblad operators of Eq.~(\ref{eq:dephasing-lindblad-x}) cause
decay of density matrix elements off-diagonal in the $\sigma^{x}$
basis and do not preserve parity. In the absence of a unitary part (ie
for $H=0$), any product state of eigenstates of $\sigma^x_j$ is a
steady state as is a statistical mixture of such states; while in the
presence of a generic, time-dependent $H$ the single steady state is
the identity matrix and the steady state is the fully mixed state
$\rho\propto\mathbb{I}$.

The right panel of Fig.~\ref{fig:Real-time-plot-x-dephasing} shows the
Fourier transform of the evolution, confirming that
\begin{itemize}
\item In both cases there is a peak at $1/2$ the driving frequency, and
\item Slower driving does result in stronger damping, broadening the
  peak of the transform.
\end{itemize}

We quantify the decay rate by fitting the quantity
$\left|\left<\sigma_{1}^{x}\right>(t)\right|$ with an exponentially
decaying function, ie, by determining $\xi_{t}$ in
$\left|\left<\sigma^{x_{1}}\right>(nT)\right|=\sigma\left(0\right)\exp(-\xi_{t}nT)$.
The results of this fit are shown in
Fig.~\ref{fig:decay-rate-plots-tau}: The left panel shows the
dependence of $\xi_{t}$ for fixed $\Gamma$ and varying $T_{z}$, while
the right panel shows the same quantity for varying $T_{z}$ but now
keeping $\Gamma T_{z}$ fixed, demonstrating that it is the quantity
$\Gamma T_{z}$ that determines the damping rate rather than $\Gamma$
or $T_z$ alone. 

A simple picture for this is provided by a minimal model for this type
of driving for a single spin presented in Appendix
\ref{sec:Analytical-approach}. There it is shown that if the flipping
(paramagnetic) part acts for time $\tau$, the Lindblad operator is
proportional to some $\gamma$ and the transverse field is $h$ such
that $h \tau = \pi$ (thus in the absence of the dephasing term would
exactly flip the spin) then the $x$ component after time $\tau$ is
$\left\langle \sigma^{x}\right\rangle
\left(\tau\right)=\mathrm{tr}\left(\rho\left(\tau\right)\sigma^{x}\right)=\rho_{x}(\tau)$
with
\begin{equation}
  \rho_{x}(\tau)=\rho_{x}\left(0\right)\exp\left(-\gamma\tau\right)\cos\left(\pi\sqrt{1-\gamma^{2}\tau^{2}/\pi^{2}}\right),
  \label{eq:rhox-one-rotation}
\end{equation}
so that
\begin{itemize}
\item the spin loses polarisation; and
\item the rate of loss of polarisation (damping rate) depends on the
  product $\gamma \tau$ rather than on each factor individually.
\end{itemize}

These properties we also find for our many-spin problem, where the
rate of damping only depends on the product $\Gamma T_z$ as shown in
the right panel of Fig.~\ref{fig:decay-rate-plots-tau}, suggesting
that the toy model of the appendix correctly describes the relevant
physics for this type of Lindblad operator. In particular this shows
that the more rapid the flipping of the spins, the longer the lifetime
of the DTC for a given damping rate $\Gamma$.

\begin{figure*}
  \includegraphics[width=1\columnwidth]{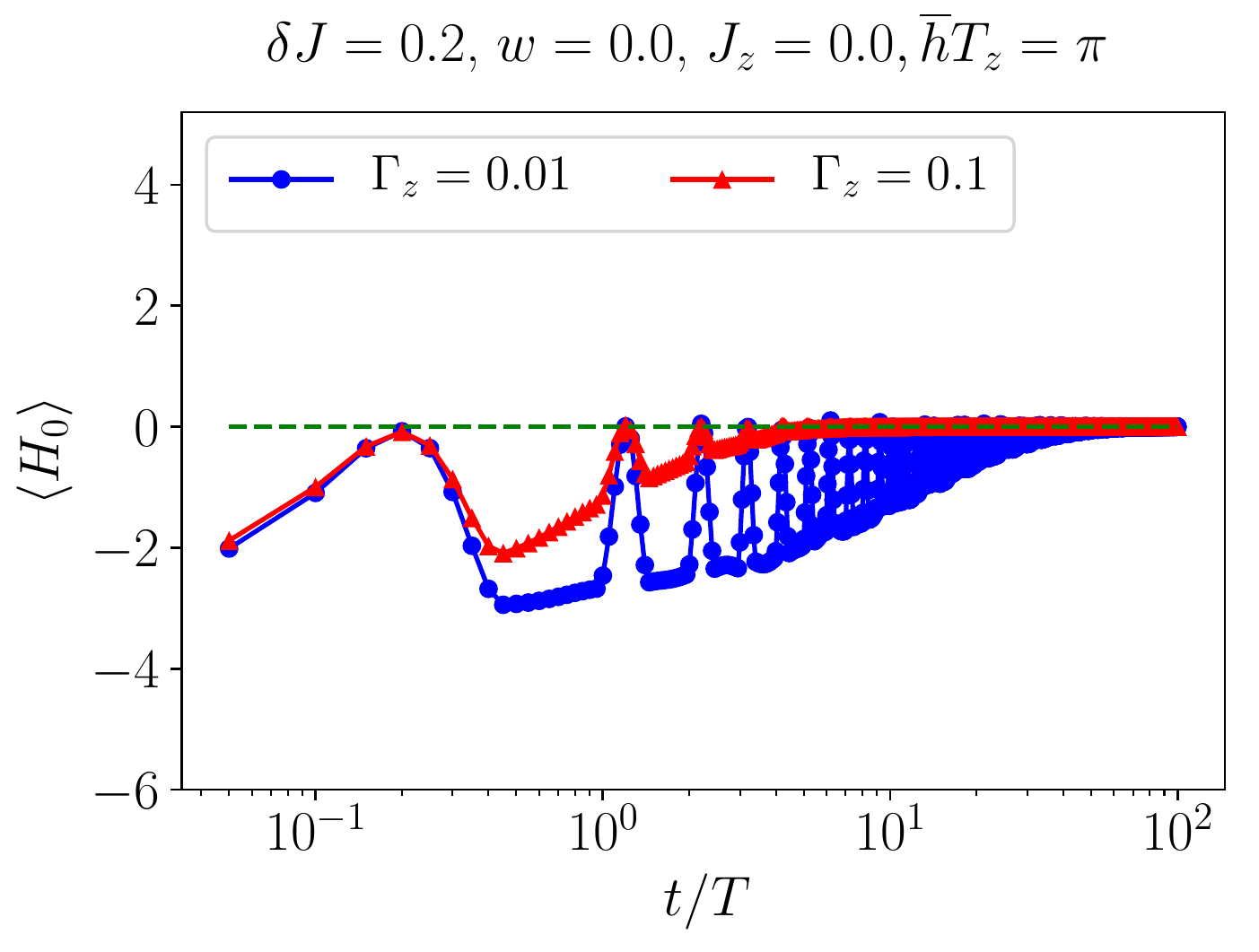}
  \includegraphics[width=1\columnwidth]{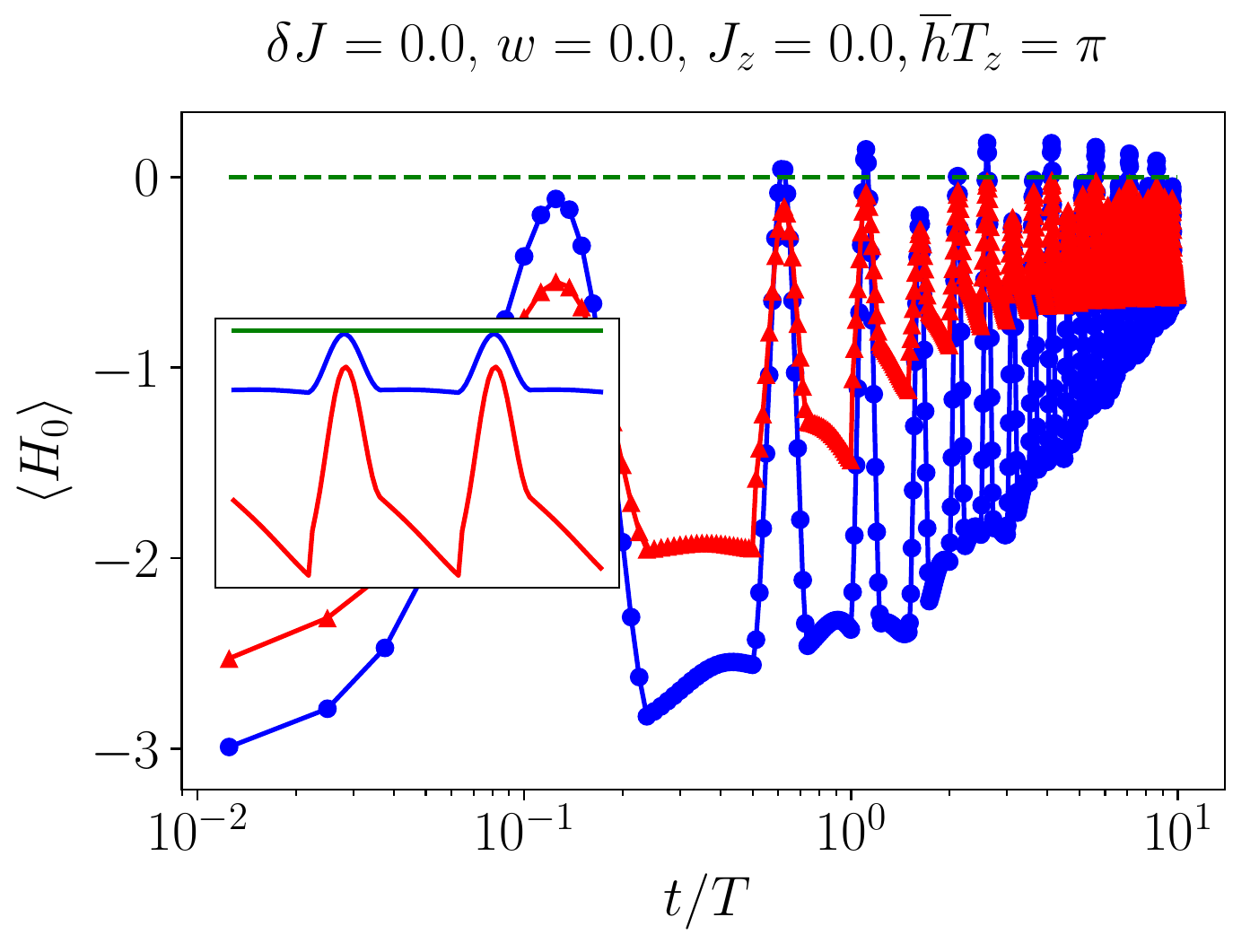}
  \caption{Instantaneous expectation value of average Hamiltonian
    $\left\langle H_{0}\right\rangle $ for the case of dephasing
    (\textbf{left} panel, see \ref{subsec:dephasing-in-z}) and
    ``thermal'' (\textbf{right} panel, see \ref{sec:fin-temp-bath})
    Lindblad operators. The green lines indicate the average of all
    the energy eigenvalues, which is the expectation value of $H_{0}$
    in the infinite-temperature Gibbs state (or the fully-mixed
    state). In the dephasing case the final state is the fully-mixed,
    featureless state. In the ``thermal'' case the steady-state is a
    time-periodic state with energy below that of the infinite
    temperature state. The blue/red trace correspond to bath inverse
    temperature $\beta=0.5$ and $0.05$, respectively. The inset shows
    the expectation value over two cycles of this steady state with a
    linear time axis (this is obtained by direct calculation at a very
    large time of $t/T\sim4000$ but the results are the same for
    larger times: this is a periodic steady-state. For dephasing
    operators in the $x$ direction the results are very similar to
    those shown in the left panel. \label{fig:energy-plots}}
\end{figure*}

\subsubsection{Parity-preserving dephasing:  $z$ direction}
\label{subsec:dephasing-in-z}

The second type of Lindblad operator we consider is
\begin{equation}
  L_{j}=\sqrt{\Gamma}\sigma_{j}^{z},
  \label{eq:dephasing-lindblad-z}
\end{equation}
causing dephasing in a direction aligned with the transverse field.
Under the action of this type of dynamics, parity is a conserved
quantity: as follows from Eq.~\ref{eq:commutation-adjoint},
$\mathrm{{tr}\left(P\rho\left(t\right)\right)}=\mathrm{{tr}\left(P\rho\left(0\right)\right)}$
for all $t$, with $P$ the parity operator of Eq.~\ref{eq:parity}.

Real time and Fourier transformed results for this are shown in
Fig.~\ref{fig:Real-time-plot-z-dephasing}. The difference from the
case of~\ref{sec:dephasing-x} is that now the decay occurs throughout
the evolution, rather than mostly during the ``flipping'' $L_z$ part,
therefore there are no longer plateaus in the expectation value of
$\sigma^x_j$. Otherwise the overall behaviour is very similar to the
previous case: The DTC is destroyed. This is a concrete demonstration
that, as follows from the arguments of \ref{sec:mixing-blocks}, it is
not enough that the symmetry is preserved for the survival of the DTC
phase.

\subsection{Finite-temperature bath}
\label{sec:fin-temp-bath}

As the operators of both Eq.~(\ref{eq:dephasing-lindblad-x}) and
Eq.~(\ref{eq:dephasing-lindblad-z}) are Hermitian, the fully-mixed
density matrix $\rho=\mathbb{I}$ is a solution. This state corresponds
to infinite temperature (see Fig.~\ref{fig:energy-plots}), and is not
unexpected since these Lindblad operators cause transitions between
states without any energy preference. To show that this is not what
destroys the DTC, we now select a set of Lindblad operators that in
the static limit would have a finite-temperature Gibbs state as the
steady-state. Under periodic driving, there are two processes at play:
the unitary evolution, which tends to increase the energy of the
system, and the dissipative evolution which tends to take the system
to some finite-energy state. While their interplay and resulting
steady-state are complicated, we find that the long-time state is not
infinite temperature but still does not display DTC.

\begin{figure*}
  \includegraphics[width=1\columnwidth]{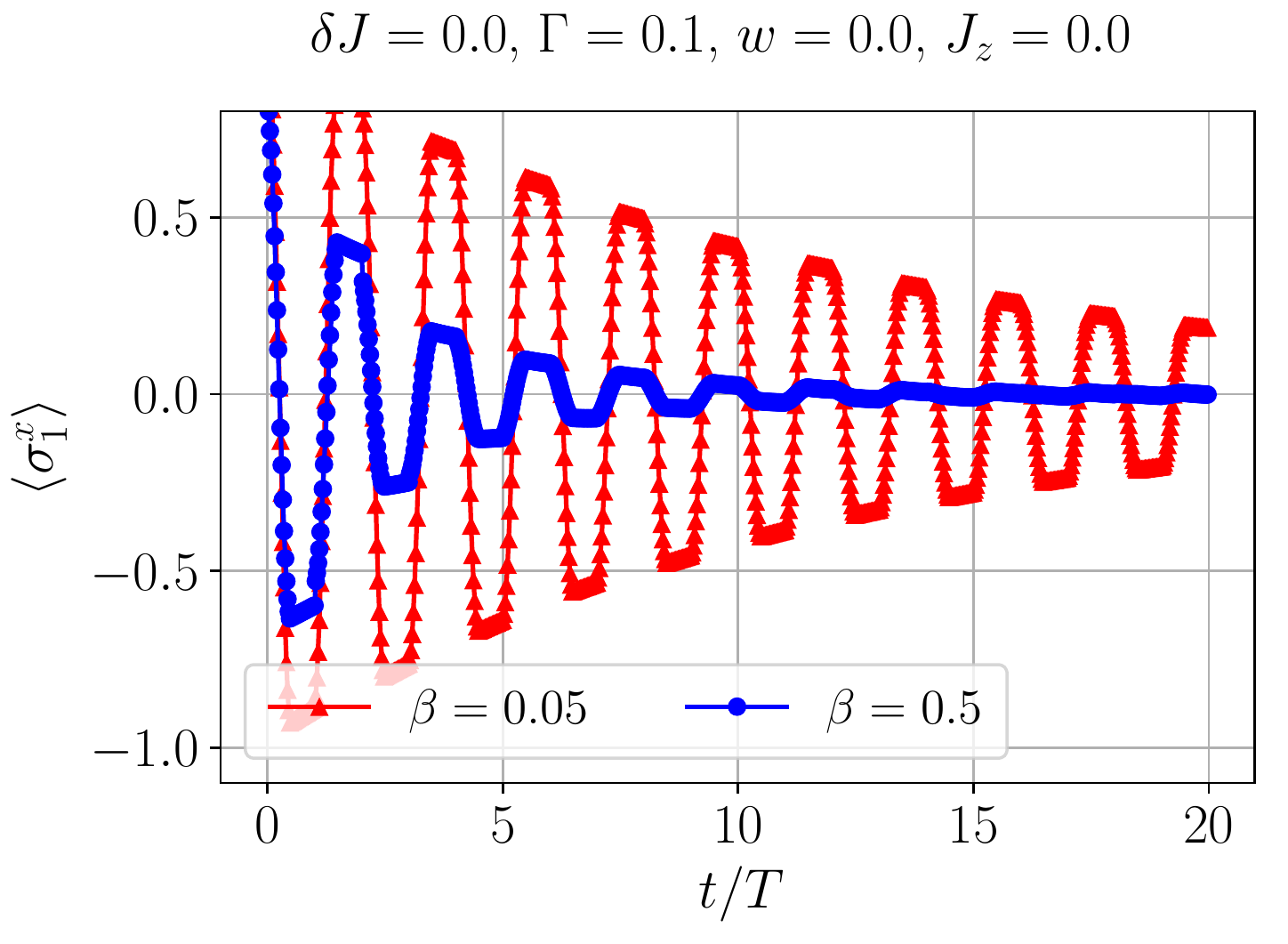}
  \includegraphics[width=1\columnwidth]{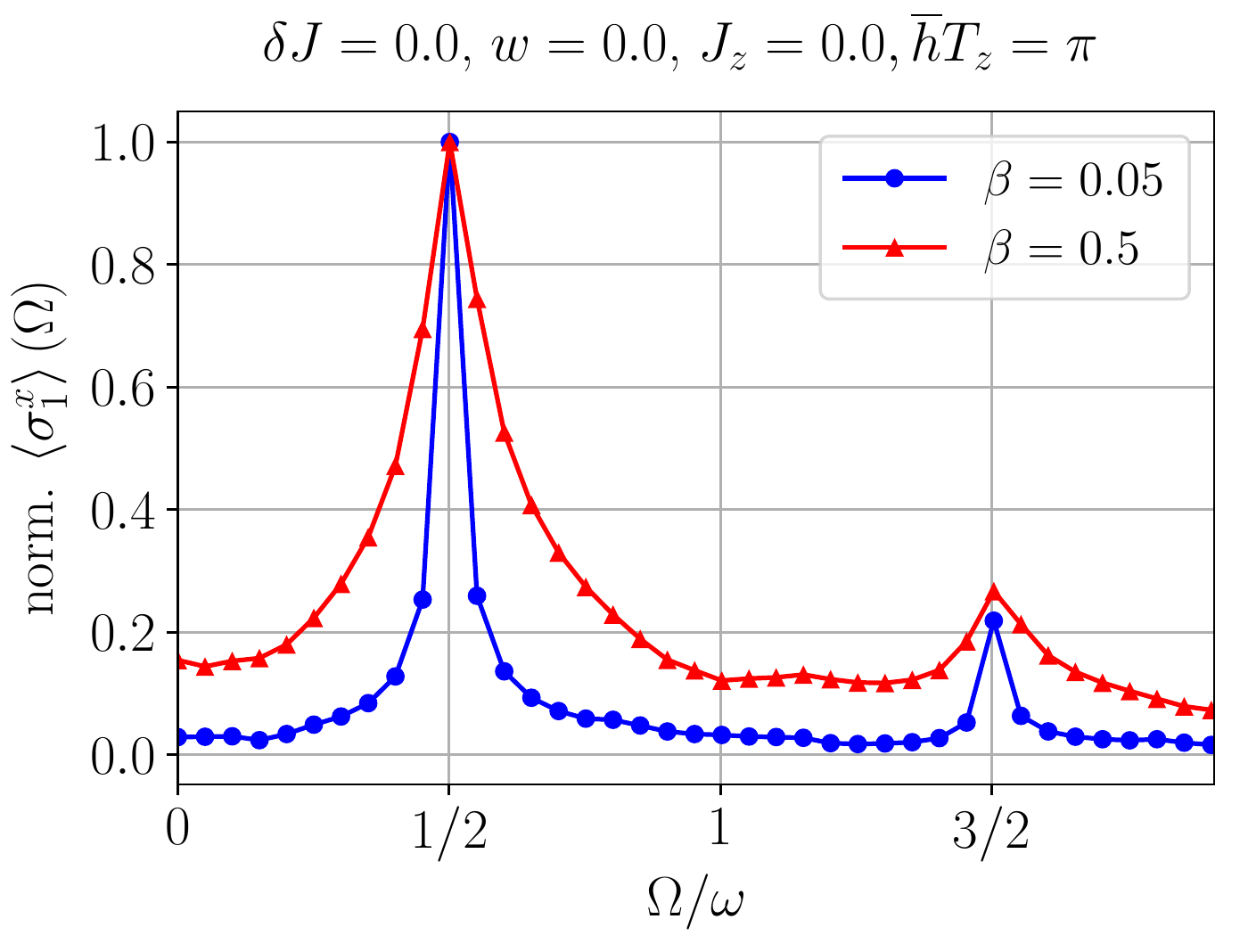}
  \caption{\textbf{Left}:
    Real time plot of
    $\left\langle \sigma_{1}^{x}\right\rangle \left(t\right)$ using
    the ``thermal'' Lindblad operators of
    Sec.~\ref{sec:fin-temp-bath}, for times
    (see Eq.~\ref{eq:hising}) $T_{x}=T_{z}=1$. Such operators lead to
    thermal (Gibbs) steady states in the static case at some inverse
    temperature $\beta$. The blue (triangular) and red (circular)
    markers indicate different environment temperatures, and the
    system-bath interaction does not couple eigenstates of different
    parity. The data displayed is for size $L=5$ and the system-bath
    coupling strength $\gamma=0.1$.
    \textbf{Right}: Fourier transform for the
    same parameters, showing the broadening due to the
    finite-temperature bath. The main insight to be gained from this
    figure is that generic system-bath couplings, even at finite
    temperatures (see Fig.~\ref{fig:energy-plots}), destroy the DTC.
    \label{fig:Real-time-plot-master-equation}}
\end{figure*}

In the static case, a thermal bath is often modelled by coupling the
system of interest to an environment with a thermal energy
distribution. This is done via a term in the system-bath Hamiltonian
$H_{SB}=V\otimes B$ with $V$ acting on the system degrees of freedom
and $B$ acting on the bath degrees of freedom. Under a set of broad
assumptions, one arrives at a Markovian master equation for the system
of the form of Eq.~\ref{eq:Lindblad} with Lindblad operators of the
form
\begin{equation}
  L_{mn}=\Gamma_{mn}\ket m\bra n\label{eq:fgr-lindblad-operator}
\end{equation}
with $\ket m$ eigenstates of the instantaneous Hamiltonian,
$H\ket m=\epsilon_{m}\ket m$ and
$\Gamma_{mn}=2\pi\left|V_{mn}\right|g\left(\epsilon_{n}\right)$.  Here
$g\left(\epsilon_{n}\right)$ models everything about the external
thermal bath and $V_{mn}=\bra mV\ket n$ is a matrix element of the
system part of the bath-system coupling operator in the instantaneous
eigenstates. If this operator commutes with the parity operator,
$\left[V,P\right]=0$ and the Hamiltonian also satisfies
$\left[H,P\right]=0$ then the matrix elements between eigenstates of
different parity vanish so that the parity a conserved
quantity. The choice
$g\left(\epsilon\right)\propto\exp\left(-\beta\epsilon\right)$ ensures
that the steady-state of this static system is the Gibbs distribution
$\rho_{S}=\sum_{n}\exp\left(-\beta\epsilon_{n}\right)\ket n\bra n$ while
the time evolution up to the steady-state depends on the specific form
of the operator $V$, that is, on the details of the system-bath
coupling.

We shall apply this approach to the Floquet case by taking the bath to
act in the way just described throughout each of the two parts of the
evolution; that is, during the $H_{x/z}$ part of the evolution, the
Lindblad operators will be taken to be of the form of
Eq.~\ref{eq:fgr-lindblad-operator} with $\ket{m}$ the eigenstates of
the Hamiltonian $H_{x/z}$. This is valid so long as the bath
equilibration timescales are much shorter than the times $T_{x/z}$
over which each part of the Hamiltonian is acting.

The particular form of the coupling operator we shall take is
\begin{equation}
  V=\frac{\Gamma}{2\pi}\sum_{j}\sigma_{j}^{z}\label{eq:fgr-v-operator}
\end{equation}
which commutes with the parity and therefore does not couple the two
sectors at all.

As already anticipated, there is a fundamental difference between this
form of Lindblad operators and the dephasing operators we looked at
earlier: this choice does not lead to a fully-mixed state but rather
to some periodic steady-state. Physically this is because this type of
bath is not energetically structureless and does not cause jumps
between all pairs of eigenstates with equal probabilities; it rather
favours a form of the density matrix that is a) diagonal in the
instantaneous energy eigenstates and b) has a Gibbs energy
distribution, that is, the preferred density matrix is a statistical
mixture of eigenstates with Gibbs weights.  In the Floquet case, the
unitary dynamics rather favours the BDE form of
Sec.~\ref{sec:Description-in-terms-of-density-matrices}. This results
in a competition between the two forms, leading to nontrivial
long-time steady states. We show this explicitly by numerically
solving the Lindblad equation and plotting the expectation value of
the instantaneous Hamiltonian in the right panel of
Fig.~\ref{fig:energy-plots}. The main figure shows the time evolution
of this energy starting from the same initial state discussed in the
previous sections and up to some late times for two different values
of the inverse temperature of the external bath, $\beta$. The inset
shows the periodic steady-state reached by the system at long
times. Evidently, for both values of $\beta$, the long-time state
is not fully mixed, is not an infinite-temperature state, and has
non-trivial dynamics, in contrast to the two dephasing cases presented
earlier.

Having established the nontriviality of the long-time steady-state, in
Fig.~\ref{fig:Real-time-plot-master-equation} we show that
nevertheless the DTC order decays under the influence of this external
parity-preserving bath. In both panels the two traces indicate
different temperatures (and the same bath-system strength coupling);
both cases results in a decay, leading to a broadening of the peak in
the Fourier transform shown in the right panel.

\section{Conclusions}

In this work we have presented an analysis of the long-time periodic
steady-state of Floquet systems with nontrivial spectral
structure. These systems include the $\pi$-spin glass, a.k.a.\
discrete time crystal, in which all states appear as members of a
doublet separated by a fixed quasienergy, leading to a subharmonic
response at a doubled period $2T$. The resulting long-time density
matrix has the form of a block diagonal ensemble (BDE)
Eq.~(\ref{eq:bde-symbols}), rather than the diagonal ensemble of
systems with no special spectral structure. Such a structure is
necessary but not sufficient for the characteristic subharmonic
oscillations that characterise DTCs; a further requirement is the
presence of spin order in the eigenstates, which in one dimension
requires the presence of disorder.

We have argued in general and supported with concrete numerical
simulations that the subharmonic response is destroyed by any
reasonable model of an environment, even if the symmetry underpinning
the spectral doublets is not destroyed. This is a direct result of its
reliance on disorder, for even static eigenstate order would be
destroyed by the effect of broad-band environmental coupling. This
holds true even if the external environment has a preferred energy
scale (temperature) such that the system goes to a nontrivial
long-time periodic steady state and preserves the symmetry, as for our
``thermal'' Lindblad operators in Sec.~\ref{sec:fin-temp-bath}, for
which a steady state at finite energy density (thus, not infinite
temperature) is obtained but nevertheless the DTC is destroyed.

This study leads to a number of interesting further topics. The first
is a detailed analysis of the dependence of these phenomena on gross
system parameters -- in particular, studying the influence of system
dimensionality is a natural follow-on from recent studies of many-body
localisation in higher dimension.\cite{Chandran:2016tu}
This question is all the more pressing as one of the recent
works studying the presence of time crystallinity was in fact
undertaken on a three-dimensional system.\cite{Choi2016} This ties
in naturally with a second major question, namely how one can hope to
stabilise time-crystalline behaviour for a long ``intermediate'' time
window, even if at asymptotically long times such order is completely
lost. In particular, can one use the coupling to an environment to
prolong, rather than curtail, the lifetime of subharmonic responses of
the isolated system? There clearly remains much scope for exciting
discoveries in both theory and experiment.

\section{Acknowledgements}
\label{sec:acknowledgements}

We thank G.~J.~Sreejith and M.~Heyl for discussions as well as A.~Das,
V.~Khemani and S.~Sondhi for collaboration on earlier work.

%


\appendix

\section{The Lindblad equation}

\label{sec:the_lindblad_equation} 

We now explain how Markovian non-unitary evolution can be described
for open quantum systems via a Lindblad master equation.

\subsection{Setup}

\label{sub:lindblad-setup}

As mentioned in the text, the Lindblad equation (\ref{eq:Lindblad}) is
the most general formulation of the differential time evolution of a
system undergoing non-unitary Markovian dynamics; the operators $L$
describe the non-unitary part of the evolution and are completely
arbitrary. Its formal solution is
\begin{equation}
\rho\left(t\right)=\exp\left(\mathcal{L}t\right)\rho\left(0\right).\label{eq:lindblad-formal-solution}
\end{equation}
One intuitive interpretation of this evolution is that the system
evolves unitarily, while at random intervals undergoing a measurement
induced by the operator $L_{\alpha}$. The result of the measurement
is discarded, leaving the system in a mixed state, and the evolution
continues. As we are taking the non-unitary evolution to be due to
the interaction with some macroscopically large external environment,
this interpretation is intuitively appealing.

In general, the Lindblad equation has at least one fixed point for
which $\partial_{t}\rho=0$; more generally, the fixed point
is the eigenvector of $\mathcal{L}$ corresponding to the eigenvalue
with 0 real part. As we shall see below, the presence of symmetries
ensures the existence of multiple fixed points.

Finally, let us note that if all the $L_{a}$ are Hermitian, $L_{a}=L_{a}^{\dagger}$
then a fixed point is the identity $\rho=\mathbb{I}$. Thus for Hermitian
Lindblad operators typically the steady state is the fully-mixed,
infinite temperature state $\rho=\mathbb{I}$.

\subsection{Symmetries\label{sub:symmetries}}

Analogously to the Heisenberg picture in unitary quantum mechanics,
one may define the adjoint Lindblad operator by its action on some
observable $A$ as follows:\cite{Ribeiro2015a} 
\[
\mathcal{L^{\mathrm{ad}}}A=i\left[H,A\right]+\sum_{a}\left(L_{a}AL_{a}^{\dagger}-\frac{1}{2}\left[L_{a}^{\dagger}L_{a},A\right]_{+}\right)
\]
The Lindblad operator and its adjoint are related by 
\[
\mathrm{tr}\left(A\left(\mathcal{L}\rho\right)\right)=\mathrm{tr}\left(\left(\mathcal{L}^{\mathrm{ad}}A\right)\rho\right)
\]
from which it follows that the time-dependent expectation value is
\[
\mathrm{tr}\left(A\exp\left(\mathcal{L}t\right)\rho\right)=\mathrm{tr}\left(\left(\exp\left(\mathcal{L}^{\mathrm{ad}}t\right)A\right)\rho\right)
\]

We now concentrate on the case where the Hamiltonian commutes with
some operator, taking for concreteness the parity operator, Eq.~\ref{eq:parity}.
Noticing that one may write the adjoint operator as 
\begin{equation}
\mathcal{L^{\mathrm{ad}}}P=i\left[H,P\right]+\frac{1}{2}\sum_{a}\left(L_{a}^{\dagger}\left[P,L_{a}\right]+\left[L_{a}^{\dagger},P\right]L_{a}\right)\label{eq:commutation-adjoint}
\end{equation}
we immediately conclude that if $\left[L_{a},P\right]$ for all $a$
then the expectation value of the operator $P$ is time independent;
it therefore remains a conserved quantity.

As the operator $P$ has two possible values, this implies that there
must be two fixed points.

\subsection{Time-dependent systems\label{subsec:Time-dependent-systems}}

The Lindblad equation is first order in time and linear in $\rho$.
This implies that its solution for a time-depenent
$\mathcal{L}\left(t\right)$ can be obtained in a very similar way to
that for the Schrodinger equation. In particular, if the Lindblad
operators $L_{a}$ are time-independent and $H\left(t\right)$ as in
Eq.~\ref{eq:hising} then the operator propagating $\rho$ over a period
$T$ is
\begin{equation}
  \mathcal{K}=\exp\left(\mathcal{L}_{x}T_{x}\right)\exp\left(\mathcal{L}_{z}T_{z}\right)\label{eq:lindblad-propagator}
\end{equation}
with
$\mathcal{L}_{\alpha}\rho=-i\left[H_{\alpha},\rho\right]+\sum_{a}\left(L_{a}\rho
  L_{a}^{\dagger}-\frac{1}{2}\left[L_{a}^{\dagger}L_{a},\rho\right]_{+}\right)$
for $\alpha=x,z$ so that time evolution is generated by
$\rho\left(T\right)=\mathcal{K}\rho\left(0\right)$.  This defines a
quantum map. We note that $\mathcal{K}$ need not be the exponential of
a time-dependent Lindblad operator.

\section{Analytical approach to dephasing in the $x$ direction}
\label{sec:Analytical-approach}

Because of the special feature of this model that decay occurs only
during the rotation phase when $H_z$ is acting, and because $H_z$ is a
sum of single-spin terms, we can write down a single-spin simplified
version of the problem which reproduces the observed dependence of the
rates on the parameters.

During the part of the period over which the ferromagnetic term
$H_{x}$ is acting and the system is evolving according to
$\mathcal{L}_{x}$, the spins align along
the $x$ direction.  While so aligned, the dephasing term has no effect
as any state with all spins (anti)aligned with the $x$ axis is a
steady state of $\mathcal{L}_{x}$ with the Lindblad operators of
Eq.~(\ref{eq:dephasing-lindblad-x}).  On the other hand, the
paramagnetic term $H_{z}$ rotates the spin at position $i$ by
$h_{i}T_{z}$ which is on average $\pi$; thus the spins approximately
flip. This flipping does not have to be exact, due to the action of
the $H_{x}$. However during the process of flipping each spin is
obviously not in a steady state of $\mathcal{L}_{z}$ so the
dephasing term is effective.

We now abstract this process into a single-spin evolving under
single-spin versions of $H_{z}$ and the Lindblad operators, taking
\begin{equation}
  H=\frac{1}{2}h\sigma^{z}\label{eq:single-spin-h}
\end{equation}
and a dephasing Lindblad operator
\begin{equation}
  L=\sqrt{\gamma}\sigma^{x}\label{eq:single-particle-lindblad}
\end{equation}
with the density matrix evolving as in Eq.~\ref{eq:Lindblad}.  We will
use this to propagate our spin forward in time by $\tau=h/\pi$, so
that in the unitary case $\gamma=0$ the spin's $x$ component would
exactly flip.

The density matrix can be decomposed as 
\[
\rho=\rho_{0}\mathbb{I}+\rho_{x}\sigma^{x}+\rho_{y}\sigma^{y}+\rho_{z}\sigma^{z}
\]
and Eq.~\ref{eq:Lindblad} results in
\begin{equation}
  \begin{array}{ccc}
  \partial_{t}\rho_{0} & = & 0\\
  \partial_{t}\rho_{x} & = & -h\rho_{y}\\
  \partial_{t}\rho_{y} & = & h\rho_{x}-2\gamma\rho_{y}\\
  \partial_{t}\rho_{z} & = & -2\gamma\rho_{z}.
  \end{array}\label{eq:dynamics-spin-components}
\end{equation}
The $\rho_{0}$ is constant, preserving the trace, while $\rho_{z}$
decays exponentially if it is not zero. The $x$ and $y$ components
perform damped oscillations, as can be seen by writing $\omega_{0}=h$
and $\eta=\gamma/h$ whence
\begin{equation}
  \partial_{t}^{2}\rho_{x}+2\eta\omega_{0}\partial_{t}\rho_{x}+\omega_{0}^{2}\rho_{x}=0\label{eq:dynamics-damped-sho}
\end{equation}
which is a damped harmonic oscillator: for $\eta>1$ or $\gamma>h$ it is overdamped
and the spin decays exponentially (without any oscillations) to
$\rho_{x}=\rho_{y}=\rho_{z}=0$. On the other
hand, for $\eta<1$ or $\gamma<h$ it is underdamped, decaying to the
same point but oscillating on the way with frequency 
$\sqrt{h^{2}-\gamma^{2}}$ and decay constant $\gamma$.

Integrating this for time $\tau=\pi/h$, which in the absence of
damping would result in the spin flipping, the condition for
being underdamped is
\[
  \tau\gamma<\pi
\]

For the initial condition $\rho_{y}=\rho_{z}=0$ and $\rho_{x}\neq0$  we obtain the result displayed in
Eq.~(\ref{eq:rhox-one-rotation}).

\end{document}